\documentclass[journal,onecolumn]{IEEEtran}

\usepackage{cite}
\usepackage{amsmath,amssymb,amsfonts}
\usepackage{graphicx,color}
\usepackage{textcomp}
\usepackage{algorithm, algpseudocode}
\usepackage{authblk}

\usepackage{svg}
\usepackage{hyperref}
\usepackage{soul}
\hypersetup{colorlinks=true,allcolors=blue}
\usepackage{paralist}
\usepackage{dirtytalk}
\usepackage{hypcap}
\usepackage{tikz}
\usepackage{graphicx}
\usepackage[commentColor=black]{algpseudocodex}
\usetikzlibrary{positioning, calc, shapes, snakes}
\usepackage{graphicx}
\usepackage{subcaption}
\usepackage{tabularx}
\usepackage{booktabs}
\usepackage{shellesc}

\algblock{Input}{EndInput}
\algnotext{EndInput}
\algblock{Output}{EndOutput}
\algnotext{EndOutput}

\def\BibTeX{{\rm B\kern-.05em{\sc i\kern-.025em b}\kern-.08em
    T\kern-.1667em\lower.7ex\hbox{E}\kern-.125emX}}
\AtBeginDocument{\definecolor{ojcolor}{cmyk}{0.93,0.59,0.15,0.02}}
\def\OJlogo{\vspace{-4pt}\hskip-4pt\includegraphics[height=18pt]{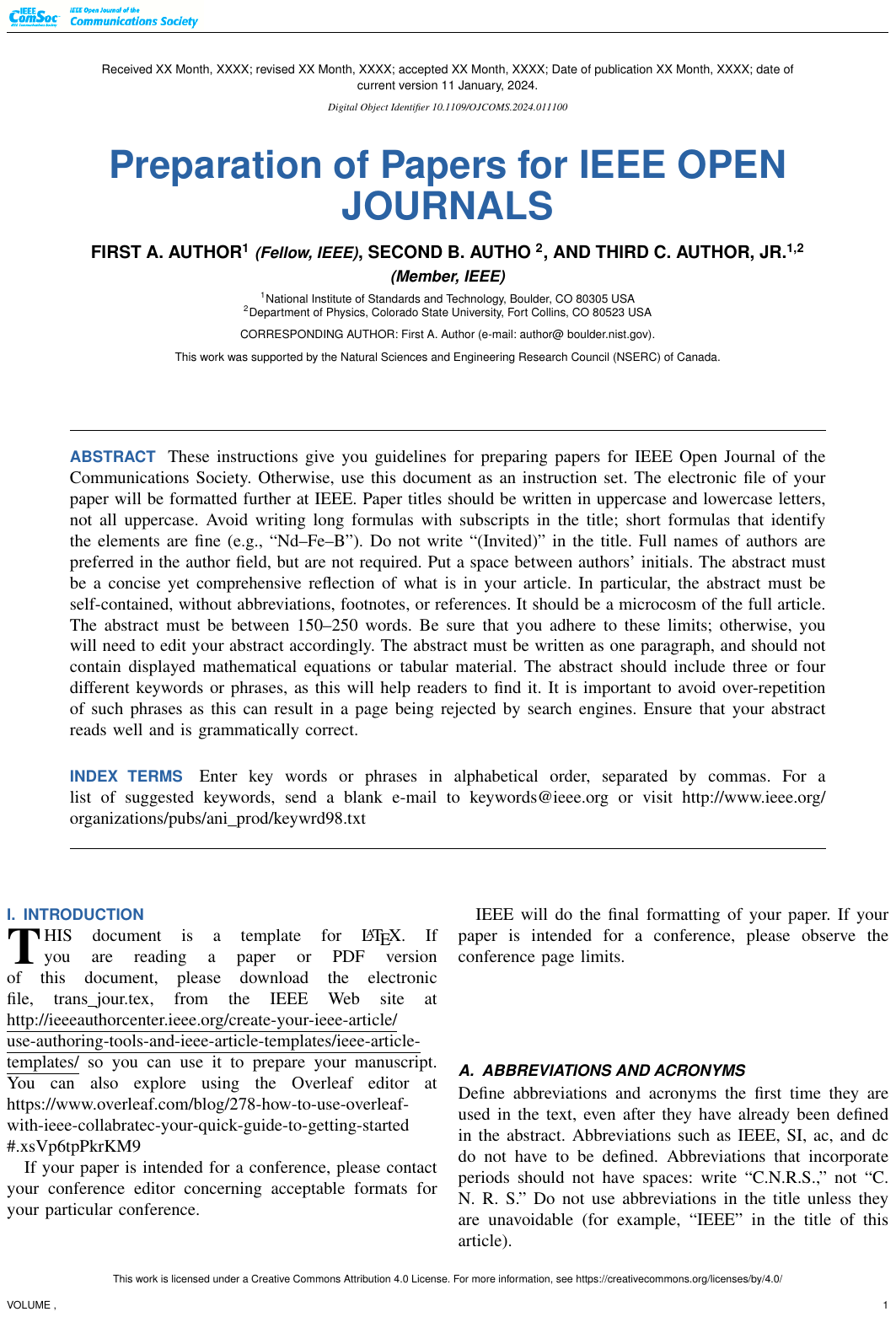}}

\algrenewcommand{\alglinenumber}[1]{\bfseries\footnotesize #1}
\algrenewcommand{\textproc}{}

\definecolor{pinegreen}{cmyk}{0.92,0,0.59,0.25}
\definecolor{royalblue}{cmyk}{1,0.50,0,0}
\definecolor{lavander}{cmyk}{0,0.48,0,0}
\definecolor{violet}{cmyk}{0.79,0.88,0,0}
\tikzstyle{cblue}=[circle, draw, thin,fill=cyan!20, scale=0.8]
\tikzstyle{qgre}=[rectangle, draw, thin,fill=green!20, scale=0.8]
\tikzstyle{rpath}=[ultra thick, red, opacity=0.4]
\tikzstyle{legend_isps}=[rectangle, rounded corners, thin, 
                       fill=gray!20, text=blue, draw]
                        
\tikzstyle{legend_overlay}=[rectangle, rounded corners, thin,
                           top color= white,bottom color=green!25,
                           minimum width=2.5cm, minimum height=0.8cm,
                           pinegreen]
\tikzstyle{legend_link}=[rectangle, rounded corners, thin,
                           top color= white,bottom color=red!25,
                           minimum width=2.5cm, minimum height=0.8cm,
                           red]
\tikzstyle{legend_ground_space_link}=[rectangle, rounded corners, thin,
                           top color= white,bottom color=red!25,
                           minimum width=2.5cm, minimum height=0.8cm,
                           red]
\tikzstyle{legend_phytop}=[rectangle, rounded corners, thin,
                          top color= white,bottom color=cyan!25,
                          minimum width=2.5cm, minimum height=0.8cm,
                          royalblue]
\tikzstyle{legend_general}=[rectangle, rounded corners, thin,
                          top color= white,bottom color=lavander!25,
                          minimum width=2.5cm, minimum height=0.8cm,
                          violet]

\begin{document}

\title{On the Role of Communications for Space Domain Awareness}


\author[1]{Nathaniel G. Gordon}
\author[2]{Nesrine Benchoubane}
\author[2]{Güneş Karabulut Kurt}
\author[1]{Gregory Falco}

\affil[1]{Cornell University, Sibley School of Mechanical and Aerospace Engineering, \{ngg29, gfalco\}@cornell.edu}
\affil[2]{Poly-Grames Reserach Center, Department of Electrical Engineering, Polytechnique Montréal, \{nesrine.benchoubane, gunes.kurt\}@polymtl.ca}




\markboth{Gordon et al.: On the Need for Secure Space Domain Awareness Systems}
{Gordon et al.: On the Need for Secure Space Domain Awareness Systems}
%



\maketitle

\begin{abstract}
\textbf{Space Domain Awareness (SDA) has become increasingly vital with the rapid growth of commercial space activities and the expansion of New Space. This paper stresses the necessity of transitioning from centralized to distributed SDA architectures. The current architecture predominantly relies on individual downhaul, which we propose to transition to on-orbit distribution. Our results demonstrate that the individual downhaul architecture does not scale efficiently with the increasing number of nodes, while on-orbit distribution offers significant improvements. By comparing the centralized architecture with the proposed distributed architecture, we highlight the advantages of enhanced coverage and resilience. Our findings show that on-orbit distribution greatly outperforms individual downhaul in terms of latency and scalability. Specifically, the latency results for on-orbit distribution are substantially lower and more consistent, even as the number of satellites increases. In addition, we address the inherent challenges associated with on-orbit distribution architecture, particularly cybersecurity concerns. We focus on link security to ensure the availability and integrity of data transmission in these advanced SDA systems. Future expectations include further refinement of on-orbit distribution strategies and the development of robust cybersecurity measures to support the scalability and resilience of SDA systems.}
\end{abstract}


%
\IEEEpeerreviewmaketitle





\section{Introduction}

\begin{figure*}[ht]
\centering
    \includegraphics[scale=0.3]{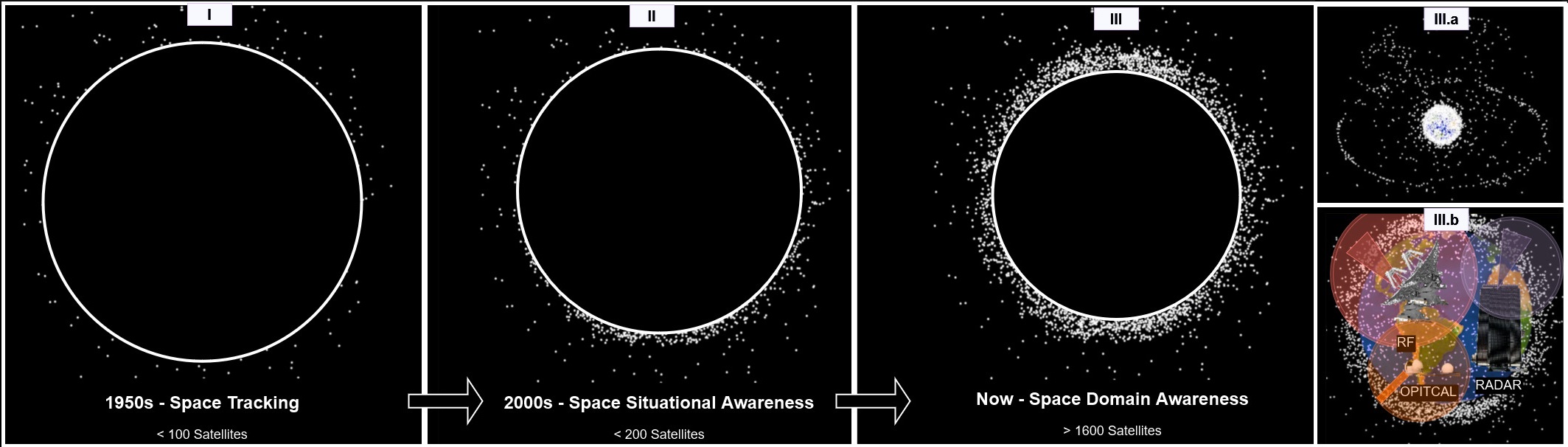}
\caption{Evolution of satellite Numbers, SDA and current ground tracking systems from 1950s to present.}
\label{fig:sda_overview}
\end{figure*}

Attempts to surveil and monitor the state of space objects date back to the 1950s\cite{borowitz20239}. Originally known as \say{space tracking}, the maintenance of tracks on specific resident space objects (RSOs), it later evolved to \say{Space Situational Awareness} (SSA), describing a larger-scale awareness of the state of the near-Earth orbital (NEO) environment. The field is now typically referred to as \say{Space Domain Awareness} (SDA), a term that encapsulates both the capture of raw data concerning RSOs and, deeper layers of characterization and meta-analysis that result from a rich network of multimodal sensing platforms \cite{blake2011space}. SDA networks are now understood as a space system because of interest for their ability to assist spacecraft operators with space traffic management, collision avoidance, managing terrestrial observations, and identifying potential space-based threats.

Over the past few decades, nations engaged in space exploration have continually striven to improve their capabilities for identifying, tracking, and analyzing objects in space. This pursuit for new tools and resources is driven by the need for SDA, which embodies the ability to detect, predict, and assess the future trajectories of objects, thereby safeguarding existing space operations in the defense domain. This imperative is reinforced by the growth of the global space economy, which has increased 2.5 times between 2005 and 2021 \cite{spacefoundation2022}.

SDA applications span a broad range, encompassing new space missions, cislunar and deep space missions and more. These missions face significant challenges, including debris management, cybersecurity threats, environmental hazards, and the complexities of coordinating various operations in space. Despite these challenges, all missions have fundamental similarities in their constitution. The components of any space system can broadly be described as falling into four main segments: 
\begin{inparaenum}[(i)]
    \item a ground segment, consisting of terrestrial networks, ground stations, telescopes, and data centers;
    \item a link segment, composed of radio-frequency and/or optical links between assets of the network;
    \item a space segment, consisting of space-based hardware;
    \item a user segment, which allows a designated set of users to access the network to acquire data or submit commands \cite{elbert2014satellite}. 
\end{inparaenum}

In the case of SDA networks, we assume that the user consists of an operating government body, responsible for collecting and disseminating SDA insights \cite{Major2014}. The specific requirements of the SDA network make its architectural needs fall within the spectrum of other space-based wireless networks\cite{celandroni2013survey}. Most notably, it is differentiated by the heterogeneous nature of data entering the network, which varies in terms of geographic position (ground- or space-based), sensing modality, and stakeholder -- should the SDA network be tasking commercial assets for data collection and analysis \cite{blake2012space}.


Acquisition of SDA data can occur both terrestrially and on multiple orbital regimes. Each orbital regime comes with a unique set of benefits and challenges when utilized in a wireless satellite network.

One of the key orbital regimes is the Low Earth Orbit (LEO) which is characterized as the regime located closest to the Earth's surface. LEO orbits may be as low as a few hundred kilometers and, extend to a maximum of 2000 km. The low cost of inserting satellites into LEO makes it an attractive target for commercial ventures, with both Eutelsat, OneWeb and Starlink, the two largest constellations with more than 600 and 6000 satellites, respectively, utilizing this regime for their operations \cite{onewebonline, nasastarlinkonline}. The adoption of LEO for many satellite applications has led to what is known as the \say{proliferated low Earth orbit} paradigm, a high volume of satellites occupying this orbit. The combination of the existing LEO infrastructure and its proximity to the Earth's surface make it an attractive choice for serving as the foothold of an SDA network. The Space Development Agency's Transfer Layer plan identifies LEO as the regime of choice for building their next generation SDA network, although there is currently no public information on the execution details of this initiative \cite{SDAonline}.

Another important orbital regime is Medium Earth Orbit (MEO), which comprises several orbits from approximately 2000 to 36 000 km. MEO orbits, particularly low ones, present similar advantages to LEO orbits, but a nascent infrastructure means that expansion of SDA networks into this domain is unlikely in the near term \cite{petrovich2023case}. 

Lastly, Geostationary Earth Orbit (GEO) is a specific orbit at 35 785 km along the Earth's equator where a satellite appears stationary from a terrestrial perspective. Orbits at this altitude and beyond (including some highly elliptical orbits) begin to experience significant latency due to electromagnetic wave propagation delay when communicating with one another or, down to, a receiver on the ground or in LEO. Analysis of RSO identification using a GEO asset produced unacceptable uncertainty, leading to favorable results for MEO assets \cite{boyd2023SDA}.  Additionally, GEO assets are limited in the quality and modality of data they can produce due to their remoteness. From a long-term perspective, as cislunar traffic increases, there will be a need to extend SDA networks to GEO and beyond \cite{knister2020evaluation, badura2023optimizing}. 

However, for the purposes of this study, we will limit our scope to exclusively ground-based and LEO infrastructure. This focus allows us to provide a detailed analysis of the most commonly used orbital regime and its interactions with terrestrial networks.

\vspace{-11pt}

\subsection{SDA Concerns}

The increasing commercialization of space is increasing congestion and the risk of collisions. More than 22 000 objects in space larger than 10 cm are monitored by the Joint Space Operations Center \cite{nationalacademies2012earth}. Another 900 000 objects are between 1 and 10 cm in size, and more than 128 million are smaller than 1 cm \cite{esa2024space}. 

From a broader perspective, sustainability is a vital concern, notably with regard to the effective management of such space debris, which can be composed of decommissioned satellites, spent rocket stages, and fragmented debris. Such debris pollutes space and jeopardizes human missions and active satellites in orbit. For instance, in 2009, 2000 new pieces of debris were caused by a collision between the Iridium 33 and Cosmos 2251 satellites \cite{Iridium}, underlining the serious consequences of collisions with space debris.

Conversely, efforts to improve sustainability in space are multifaceted and drive a wide range of debris removal research, including Sercel's work on active debris removal and cislunar space awareness. The development of Sutter algorithms \cite{sercel2024sutter} aims to facilitate space debris management and improve situational awareness in space. At the same time, Preston's work \cite{preston2024multi}on multi phenomenological space situational awareness sensors and real-time monitoring underlines the need to develop advanced multidetection strategies to cope with the rapid increase in space traffic. Similarly, research \cite{mueller2024savant} into orbital analysis and space situational awareness illustrates the complexity of managing space debris and ensuring the sustainable use of space.

To address these growing complexities, numerous innovative methods are being developed.  For example, work \cite{agia2024modeling} on optimization-based planning and control for autonomous spacecraft highlights the importance of autonomous systems in maintaining space sustainability and debris management. Complementing this, the development of interferometric trackers \cite{hutchin2024interferometric} for the detection and tracking of GTO and GEO assets, as well as contributions to catalog processing operations (C3PO) \cite{poole2024c3po}, demonstrate technological advances in space debris tracking and cataloging objects.

In addition to operational challenges, the growth of cybersecurity threats adds another layer of complexity to SDA.  As space assets become more interconnected and reliant on data exchange, they become potential targets for cyberattacks. In February 2022, SpaceX provided Starlink terminals to Ukraine to maintain internet access during the Russian invasion \cite{Laursen2023}. Soon after their deployment, Starlink's service was disrupted by signal jamming, an electronic warfare tactic aimed at the frequency band used by Starlink. This incident, among others, highlights growing cybersecurity concerns regarding critical space infrastructure, even more so when attached to a national security scope \cite{boschetti2022space}.

Finally, we introduce Figure \ref{fig:sda_overview}
to comprehensively sum up the dramatic evolution of satellite numbers from the 1950s to the present day, as seen through a representation of eclipsing Earth. This visual representation highlights the progression from early space tracking efforts to modern SSA and SDA systems. The figure is divided into three distinct time periods:
\begin{itemize}
    \item  Figure \ref{fig:sda_overview}.I (1950s): Depicts the initial phase of space exploration, characterized by a few early satellites;
    \item Figure \ref{fig:sda_overview}.II (2000s): Shows the substantial increase in satellite numbers;
    \item Figure III (Current Date): Further subdivided into two views:
    \begin{itemize}
        \item Figure \ref{fig:sda_overview}.III.a: A geosynchronous view \cite{NASApic};
        \item Figure \ref{fig:sda_overview}.III.b: A LEO view perspective \cite{NASApic}, showcasing radar, optical, and radio frequency (RF) tracking technologies used to manage the increasingly congested LEO environment.
    \end{itemize}
\end{itemize}


\subsection{Prior Art}

The improvement of SDA network capabilities is the subject of multiple ongoing research agendas. Several efforts have been made to develop and incorporate new sensing modalities into the SDA landscape \cite{tommila2024mission, jaunzemis2016evidence,vasso2021optimal}. Expanding the variety of sensing modalities enables more nuanced downstream analyses and improves the resilience to flaws in any one sensor or sensing modality involved in with an observation. The efficient tasking of data collection, routing of data through the network and, fusion of multiple data sources presents another avenue for improving the state of the art for SDA analysis \cite{bisio2007satellite, long2014satellite,cardin2021snare}. 

In addition, there is a push to develop more robust classification systems to support the analysis of raw data products, such as the debris ontology for resident space objects \cite{hart2016new}. Once insights are classified, there is a further need for effective cognition and actuation of SDA from decision makers, be they fully autonomous or human-in-the-loop systems \cite{holzinger2018challenges}.

From another perspective, SDA networks can serve as the backbone for a commodified market of data captured from space-based assets. There has been significant investigation to show the viability of such a market \cite{wozniak2023coordinated, spacewatchonline, 10535109}. Attempts to quantify performance have focused on network coverage as a prime performance metric \cite{vasso2022multi}. In this work, however, we contend that a broader set of metrics must validate SDA network architectures. 

This brings us to examine the architecture of SDA systems and how they are deployed to effectively track assets and prepare for cyberattacks. Currently, most SDA systems report to a centralized data repository and are thus susceptible to a single point of failure. However, our work demonstrates the need for a transition to distributed architectures. 

These distributed systems are better equipped to handle the increasing complexity and scale of modern space activities. They offer enhanced coverage, resilience, as demonstrated in the work of \cite{tommila2024mission}, and adaptability, aligning with the trend toward robust and heterogeneous space architectures. This approach also complements ongoing initiatives to develop Heterogeneous Space Architectures (HSA) \cite{boschetti2023hybrid} and the Defense Advanced Research Projects Agency (DARPA)'s efforts to create a reconfigurable, multiprotocol inter-satellite optical communications terminal \cite{kuperman2022space, ramakrishnan2023development} that would allow heterogeneous constellations to be connected across diverse systems. In addition, the legacy of efforts like the DARPA program System F6 to promote fractionated space systems that can deliver on multiple synergistic objectives \cite{brown2009value}. Such a system would be able to execute SDA data collection as a secondary mission objective, allowing rapid expansion of the frontiers of an integrated SDA network.

Our research aims to validate the necessity of distributed SDA architectures and propose a comprehensive framework to support their implementation. This approach not only follows the trend of distributed systems but also ensures the robustness and security of space operations, contributing to the development of a resilient and secure space services ecosystem.

In synthesis, our contributions stand at:

\begin{itemize}
    \item We underscore the critical role of SDA in the context of growing commercial space activities and satellite proliferation;
    \item We define and formalize the components of effective SDA networks and the two reference architectures, along with a set of metrics to assess their performance;
    \item We delve into the advantages and challenges associated with these architectures, including a detailed quantitative analysis of latency performance;
    \item We stress the critical importance of addressing security, scalability and robustness concerns, necessitating future studies of SDA network architectures;
    \item We propose open-source research directions with regard to standardization and technical advancements.
\end{itemize}

The paper is organized hierarchically as follows: Section \ref{SDA} discusses the growing need for better SDA in response to the growing trends in commercial and civil space activities. Section \ref{SDA Networks} presents the two reference architectures and their components. Section \ref{challenges} identifies and analyzes the specific challenges associated with implementing the networks under study. Section \ref{sec} elaborates on the cybersecurity requirements necessary to secure data transmission in these systems. Section \ref{Future Research} outlines future research directions. Finally, Section \ref{Conclusion} concludes the paper.

\section{SDA for a New Space}
\label{SDA}

The emergence of New Space led to an influx of commercial companies, reduced launch costs, and expanded space access.  This has resulted in the Earth's orbit becoming increasingly crowded. More than 9241 spacecraft are active in total today, compared with 6718 at the end of 2022, according to the Union of Concerned Scientists \cite{ucs2024satellite}. Ten years ago, the number was just over 1000. Companies such as SpaceX, Amazon, Samsung, and OneWeb are swiftly expanding their satellite constellations, featuring up to hundreds of sensor-equipped satellites due to advancements in satellite technology and miniaturization. One of the leading companies, SpaceX, has already launched more than 6,000 satellites into LEO as of March 2024 and, is forecasting up to 12 000 satellites for its Starlink system in the coming years \cite{mcdowell2024starlink}.

This expansion of satellite constellations directly impacts the scalability and robustness requirements of SDA systems. Constellations like Starlink and OneWeb, as they grow, exponentially increase the number of satellites in orbit, presenting several interconnected challenges that necessitate adaptive SDA infrastructure.

First, the sheer volume of satellites in a constellation intensifies the complexity of space traffic management. SDA systems must track not only individual satellites but also their interactions, trajectories, and collision risks. Dynamic satellite movements within a constellation demand real-time data processing and analysis capabilities in SDA systems for timely decision-making and proactive risk mitigation.

Second, the diverse mission objectives within satellite constellations amplify the data requirements for SDA systems. Satellites in these constellations serve various purposes, such as global internet coverage, Earth observation, communications, and scientific research. Each mission objective generates specific data streams that SDA systems must process, analyze, and act upon. Scalability in SDA infrastructure is crucial to handle this diversity of data sources and ensure comprehensive situational awareness across different mission contexts.

Furthermore, the evolving challenges in space operations, including space debris management, orbital congestion, and regulatory compliance, necessitate robustness in SDA systems. A robust SDA infrastructure \textit{must} adapt to changing operational environments, integrate new data sources and technologies, and withstand potential disruptions or cyber threats. This adaptability and resilience are essential for maintaining operational continuity and enhancing the overall effectiveness of space domain awareness capabilities.

\section{SDA Networks}
\label{SDA Networks}

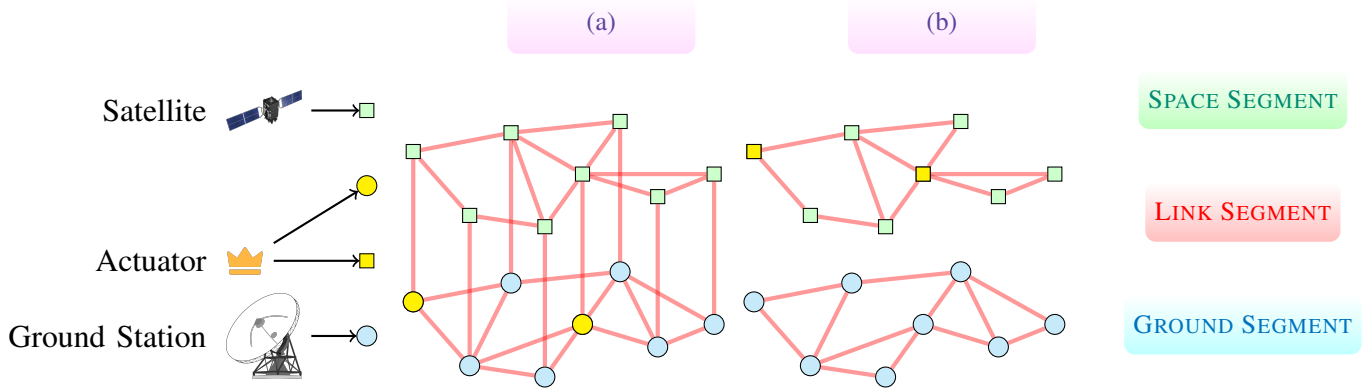
\begin{figure*}[ht]
\raggedright
    \begin{subfigure}[b]{0.3\textwidth}
    \centering
    \begin{tikzpicture}[auto, thick]
    \node[anchor=east] (gndlegend) at (-2,1) {{\large Ground Station}};
      \node[anchor=west] (ground) at (-2,1) {\includegraphics[width=1cm]{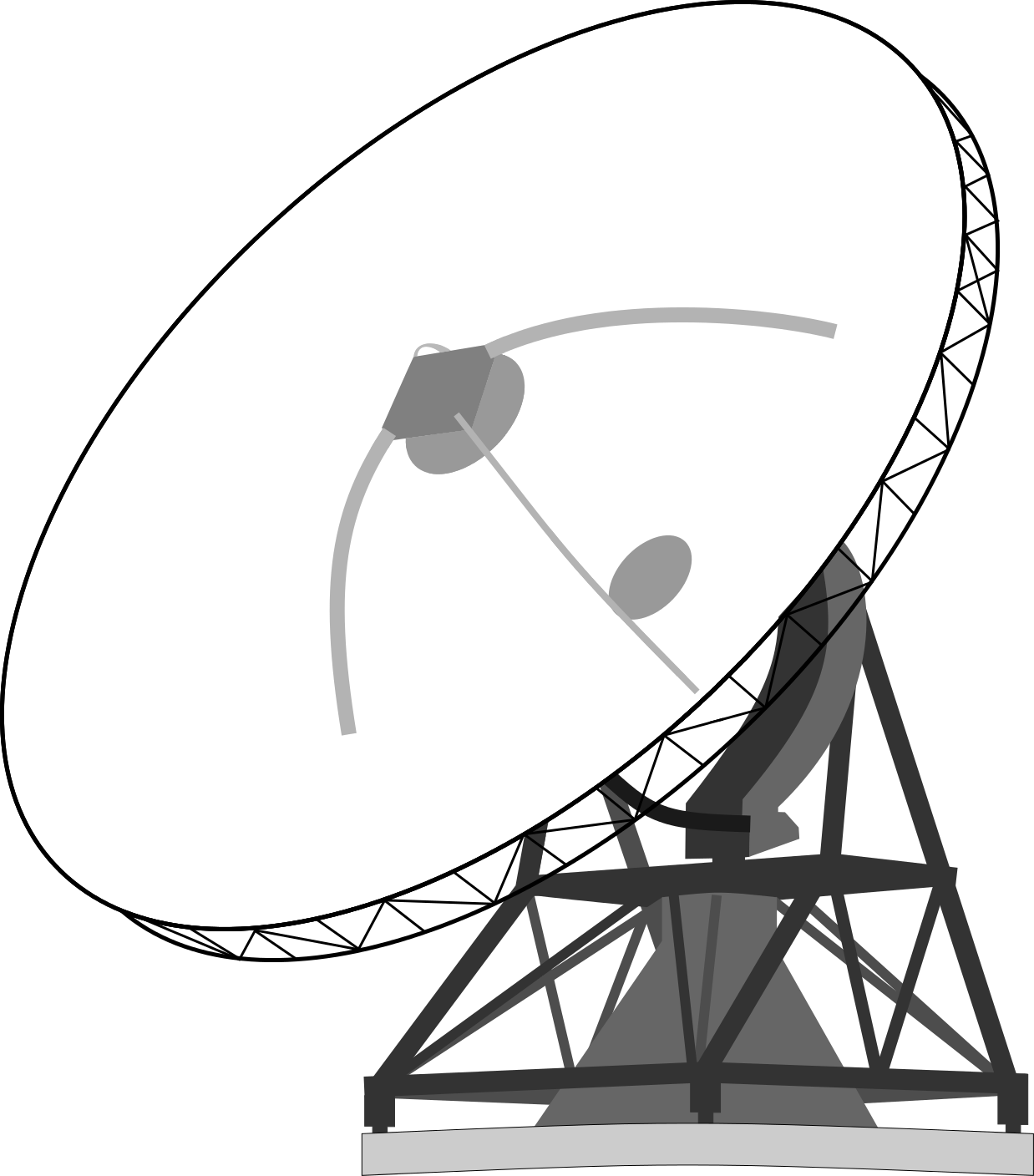}};
    \node[cblue] (cnode) at (0,1) {};
    \draw[->] (ground) -- (cnode);

    \node[anchor=east] (satlegend) at (-2,4) {{\large Satellite}};
  \node[anchor=west] (sat) at (-2,4) {\includegraphics[width=1cm]{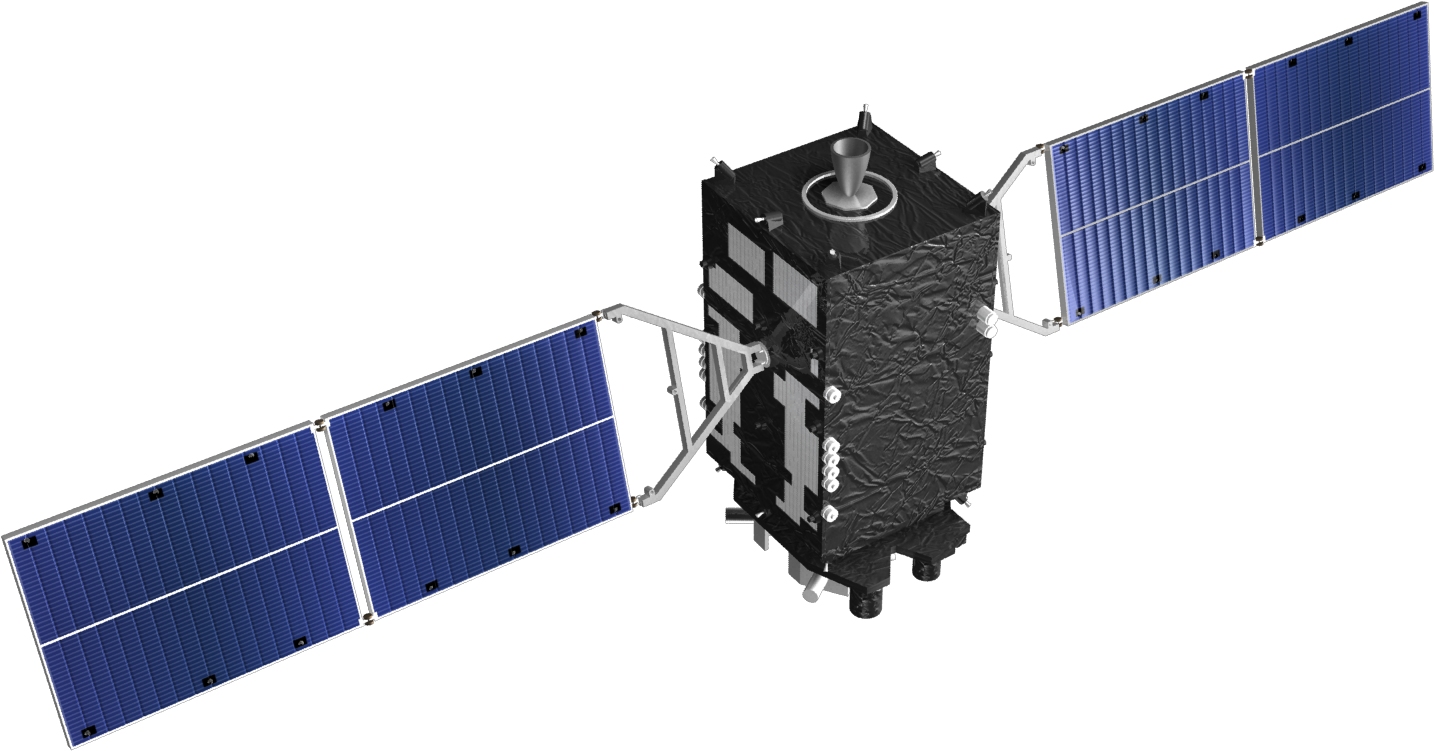}};
    \node[qgre] (qgre) at (0,4) {};
    \draw[->] (sat) -- (qgre);

    \node[anchor=east] (actlegend) at (-2,2) {{\large Actuator}};
      \node[anchor=west] (act) at (-2,2) {\includegraphics[width=.5cm]{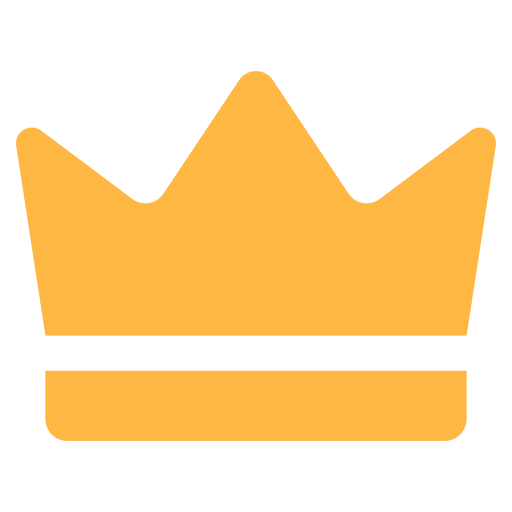}};
    \node[qgre, fill=yellow] (qgre) at (0,2) {};
    \node[cblue, fill=yellow] (cblue) at (0,3) {};
    \draw[->] (act) -- (qgre);
    \draw[->] (act) -- (cblue);
    
    \end{tikzpicture}
    \end{subfigure}%
    \begin{subfigure}[b]{0.25\textwidth}
    \centering
    \begin{tikzpicture}[auto, thick]
    
      \foreach \place/\x in {{(-2.5,0.3)/1}, {(-1.75,-0.55)/2},{(-1.2,0.55)/3},
        {(-0.75,-0.7)/4}, {(-0.25,0)/5}, {(0.25,0.7)/6}, {(0.75,-0.3)/7}, 
        {(1.5,0)/8}}
      \node[cblue] (a\x) at \place {};
     
     \path[rpath] (a1) edge (a2);
     \path[rpath] (a1) edge (a3);
     \path[rpath] (a2) edge (a3);
     \path[rpath] (a3) edge (a6);
     \path[rpath] (a2) edge (a4);
     \path[rpath] (a5) edge (a6);
     \path[rpath] (a5) edge (a4);
     \path[rpath] (a5) edge (a2);
     \path[rpath] (a5) edge (a7);
     \path[rpath] (a6) edge (a7);
     \path[rpath] (a6) edge (a8);
     \path[rpath] (a7) edge (a8);
     
    
      \foreach \place/\i in {{(-2.5,2.3)/1},{(-1.75,1.45)/2},{(-1.2,2.55)/3},
        {(-0.75,1.3)/4}, {(-0.25,2)/5}, {(0.25,2.7)/6}, {(0.75,1.7)/7},
        {(1.5,2)/8}}
        \node[qgre] (b\i) at \place {};

       \path[rpath] (b1) edge (b2);
         \path[rpath] (b2) edge (b4);
         \path[rpath] (b4) edge (b5);
         \path[rpath] (b5) edge (b7);
         \path[rpath] (b4) edge (b3);
         \path[rpath] (b7) edge (b8);
         \path[rpath] (b6) edge (b3);
         \path[rpath] (b3) edge (b1);
         \path[rpath] (b5) edge (b8);
         \path[rpath] (b5) edge (b3);
         \path[rpath] (b5) edge (b6);
          
      \foreach \i in {1,...,8}
        \path[rpath] (a\i) edge (b\i);
        \node[cblue, fill=yellow] (a1) at (-2.5,0.3) {};
        \node[cblue, fill=yellow] (a6) at (-0.25,0) {}; 
      \node[legend_general] at (0,4) {(a)};
    \end{tikzpicture}
    \end{subfigure}
    \begin{subfigure}[b]{0.3\textwidth}
        \centering
        \begin{tikzpicture}[auto, thick]
            
          \foreach \place/\x in {{(-2.5,0.3)/1}, {(-1.75,-0.55)/2},{(-1.2,0.55)/3},
            {(-0.75,-0.7)/4}, {(-0.25,0)/5}, {(0.25,0.7)/6}, {(0.75,-0.3)/7}, 
            {(1.5,0)/8}}
          \node[cblue] (a\x) at \place {};
         
         \path[rpath] (a1) edge (a2);
         \path[rpath] (a1) edge (a3);
         \path[rpath] (a2) edge (a3);
         \path[rpath] (a3) edge (a6);
         \path[rpath] (a2) edge (a4);
         \path[rpath] (a5) edge (a6);
         \path[rpath] (a5) edge (a4);
         \path[rpath] (a5) edge (a2);
         \path[rpath] (a5) edge (a7);
         \path[rpath] (a6) edge (a7);
         \path[rpath] (a6) edge (a8);
         \path[rpath] (a7) edge (a8);
         
        
          \foreach \place/\i in {{(-2.5,2.3)/1},{(-1.75,1.45)/2},{(-1.2,2.55)/3},
            {(-0.75,1.3)/4}, {(-0.25,2)/5}, {(0.25,2.7)/6}, {(0.75,1.7)/7},
            {(1.5,2)/8}}
            \node[qgre] (b\i) at \place {};
         
         \path[rpath] (b1) edge (b2);
         \path[rpath] (b2) edge (b4);
         \path[rpath] (b4) edge (b5);
         \path[rpath] (b5) edge (b7);
         \path[rpath] (b4) edge (b3);
         \path[rpath] (b7) edge (b8);
         \path[rpath] (b6) edge (b3);
         \path[rpath] (b3) edge (b1);
         \path[rpath] (b5) edge (b8);
         \path[rpath] (b5) edge (b3);
         \path[rpath] (b5) edge (b6);
         
         
          \node[legend_general] at (0,4){(b)};
          \node[legend_overlay] at (4,3){\textsc{Space Segment}};
          \node[legend_link] at (4,1.5){\textsc{Link Segment}};
          \node[legend_phytop] at (4,0){\textsc{Ground Segment}};
        \node[qgre, fill=yellow] (b1) at (-2.5,2.3) {};
        \node[qgre, fill=yellow] (b6) at (-0.25,2) {}; 
        \end{tikzpicture} 
    \end{subfigure}%
\vspace{1cm}
\caption{Comparison of reference architectures: (a) Individual Downhaul and (b) On-Orbit Distribution}
\label{fig:networks}
\end{figure*}

SDA networks must facilitate the capture, storage, and dissemination of relevant SDA data; in doing so, they must possess traits of scalability and robustness. Scalability describes the network's capacity to maintain a high degree of performance as the network's size expands. This is a significant concern to on-orbit SDA networks due to the projected increases in LEO satellite launches in the coming decade. A scalable network retains cost efficiency, low transmission latency, and data consistency even when new nodes are added to the network. 

A robust SDA network architecture is adaptable and resilient. This necessitates architecture designed with inclemency in mind, be it attrition of both ground and orbital assets, or even adversarial activity on the network. Indeed, the sensitivity of SDA data to intelligence matters makes the inherent security of the architecture a relevant concern for overall success.

\subsection{SDA Network Components}
An understanding of the key components and requirements that an SDA network sustains is necessary to comprehend how the network will perform under strain. The following is a list of the components discussed in the following work:

\begin{itemize}
    \item \textbf{Network Node:} A node on the network is a point through which data can flow. Note that a node is simply a conduit for data and, may not necessarily, perform action on it. Increasing the number of nodes on the network not only increases resilience to isolated link failures but also increases the amount of attack surface exposed to adversaries.
    \item \textbf{Data Source Node:} A data source is a sensor in the field capable of generating raw data for the SDA network. These can be satellite-mounted sensors, terrestrial telescopes, or other observation equipment.
    \item \textbf{Actuator Node:} An actuator is a node specifically granted the ability to collate SDA data and actuate on it. This could take the form of a ground-based missions operation center or an autonomous decision-making platform.
\end{itemize}

Adjusting both the distribution of nodes on the network (both in terms of geographic placement and by node type), we can form general reference architectures that could support the useful collection and actuation of SDA data.

\subsection{SDA Network Reference Architectures}

Two primary reference architectures, the Individual Downhaul and On-Orbit Distribution, offer distinct approaches to organizing and managing SDA data flows. Details of each architecture are provided below:

\subsubsection{Individual Downhaul Architecture}

Figure \ref{fig:networks}(a) represents the simplest case of how an SDA network can be implemented on existing orbital and terrestrial sensing infrastructure, outlined in operation as follows:

\begin{itemize}
    \item A single, centralized actuator node resides at a designated location on the ground;
    \item A set of satellites, ideally with heterogeneous sensing modalities and capabilities, forms the set of data source nodes;
    \item To limit additional expense, the data source nodes do not need to be capable of cross-linking. Instead, a limited number of emplacements (ground stations) serve as network nodes capable of routing SDA data downhaul from data source nodes across terrestrial networks to the actuator.
\end{itemize}

In essence, this is a unified implementation of how SDA data is sourced today (tasking a single satellite to downhaul to a ground station).

\subsubsection{On-Orbit Distribution Architecture}

Figure \ref{fig:networks}(b) limits the exposure of the attack surfaces by removing the use of the ground segment entirely. Instead, a constellation of satellites serves as the entire set of network nodes, with limited subsets serving as data sources and actuators. This provides a large set of possible permutations for implementation; the network can either all feed to a single actuator or multiple actuators. It's important to note that varying the number of network nodes present may impact the performance of the network as well as its resilience to attacks.

In the following sections, we discuss likely attack scenarios and examine the potential consequences.

\subsection{Network Performance Evaluation}

For a robust SDA system, two critical considerations emerge: Routing and Line of Sight (LOS). These factors are fundamental in assessing network performance and overall system efficiency. Routing decisions are pivotal as they determine the path data takes from its source to its destination, directly impacting the data transmission and system responsiveness. This becomes especially crucial in the context of SDA data collection, where time-sensitive information is paramount, making efficient routing practices even more critical.

We have opted to use latency as a proxy for network performance, strategically considering potential delays resulting from network rerouting and the inherent time-sensitivity of SDA data collection. Latency will serve as a valuable metric for evaluating the effectiveness and efficiency of data transmission paths within the SDA system for both architectures, and will be used in conjunction with other metrics for comparative analysis later on. In the forthcoming section, we elaborate on the algorithms employed for latency calculations.


\begin{algorithm}[!ht]
    \caption{Latency Calculation: Individual Downhaul Procedure}
    \label{alg:centralized}
    \begin{algorithmic}[1]
        \Require
        
        ${S}$, a set of $N_{total}$ satellites with Earth-centric coordinates $(p_1, p_2, p_3)$
        
        ${G}$, a set of $G_{total}$ ground stations with Earth-centric coordinates $(p_1, p_2, p_3)$
        
        ${D}$, data terminus location with Earth-centric coordinates $(p_1, p_2, p_3)$
        
        \Ensure 
        
        ${L}$, a list of latencies for each satellite in ${S}$
        
        \State ${S} \gets \{\}$
        \For{$S_i$ in ${S}$}
            \State $G_{visible} \gets \{\}$
            \For{$G_i$ in ${G}$}
                \If {has line of sight ($S_i, G_i$)}
                    \State $G_{visible} \gets G_{visible} + \{G_i\}$
                \EndIf
            \EndFor
            \If{$\ G_{visible} = \{\}$}
                \State $G_{closest} \gets (G \in G_{visible} \vert \min(\text{distance}(S_i, G)))$
                \State $T_{space} \gets \text{distance}(S_i, G_{closest}) / c$
                \State $T_{ground} \gets \text{distance}(G_{closest}, D) / c$
                \State $\mathcal{L} \gets \mathcal{L} + \{T_{space} + T_{ground}\}$
            \Else 
                \State \State $\mathcal{L} \gets \mathcal{L} + \{\}$
            \EndIf
        \EndFor 
        \For{$L_i$ in $\mathcal{L}$}
            \If{$L_i = \{\}$}
                \For {$S_j$ in ${S}$, $i \neq j$}
                    \If {has line of sight ($S_i, S_j$)}
                        \State $S_{visible} \gets S_{visible} + \{S_j\}$
                    \EndIf
                \EndFor
                \State $L_i \gets (S_j \in S_{visible} \vert \min(\text{distance}(S_i, S_j)+L_j))$
            \EndIf
        \EndFor
    \end{algorithmic}
\end{algorithm}




\begin{algorithm}[!ht]
\caption{Latency Calculation: On-Orbit Distribution.}\label{alg:distributed}
\begin{algorithmic}[1]
\Require

${S}$, a set of $N_{total}$ satellites comprised of tuples of Earth-centric coordinates $(p_1, p_2, p_3)$

${D}$, a set of $D_{total}$ deciding satellites comprised of tuples of Earth-centric coordinates $(p_1, p_2, p_3)$

\Ensure 

$\mathcal{L}$, a list of latencies for each satellite in ${S}$

\State ${S} \gets \{\}$
\For{$S_i$ in ${S}$}
    \State $D_{visible} \gets \{\}$
    \For{$D_i$ in ${D}$}
        \If {has line of sight ($S_i, D_i$)}
        \State $D_{visible} \gets D_{visible} + \{D_i\}$
        \EndIf
    \EndFor
    \If{$D_{visible} = \{\}$}
        \State $D_{closest} \gets (D \in D_{visible} \vert \min(\text{distance}(S_i, D)))$
        \State $T \gets \text{distance}(S_i, G_{closest}) / c$
        \State $\mathcal{L} \gets \mathcal{L} + \{T\}$
    \Else
        \State $\mathcal{L} \gets \mathcal{L} + \{\}$
    \EndIf
\EndFor 
\For{$L_i$ in $\mathcal{L}$}
    \If{$L_i = \{\}$}
        \For {$S_j$ in ${S}$, $i \neq j$}
            \If {has line of sight ($S_i, S_j$)}
            \State $S_{visible} \gets S_{visible} + \{S_j\}$
            \EndIf
        \EndFor
        \State $L_i \gets (S_j \in S_{visible} \vert \min(\text{distance}(S_i, S_j)+L_j))$
    \EndIf
\EndFor
\end{algorithmic}
\end{algorithm}




\section{Simulations and Analysis of SDA Networks}
\label{challenges}

Our first aim was to compare the Individual Downhaul and On-Orbit Distribution architectures under similar scenarios. This was done through a simulation of each network’s architecture. The spread of orbital nodes was simulated using the positional data of active OneWeb and Starlink assets sourced from CelesTrak \cite{celestrakonline}. A ground station network was constructed using the locations of the 12 ground installations present on the USGS International Ground Station (IGS) Network, as reported by USGS \cite{usgsigsnetworkonline}.

In the first trial, the Individual Downhaul reference architecture was simulated following the procedure outlined in algorithm \ref{alg:centralized}. The measure of a SDA data point's latency begins at a given data source node in the network. These data are first transmitted to the nearest available ground station on the network. Visibility is determined using the WGS 84 ellipsoid model of the Earth \cite{slater1998wgs} and both satellite and ground station ephemera. The latency is determined solely from the electromagnetic wave propagation time, and assumes no delay at the ground station. From the ground station, the data is transferred over fiber-optic networks across the Earth's surface to a centralized data sink. The data path is assumed to be minimized along the Earth's surface and latency is equivalent, once again, to the propagation delay. The orbital and terrestrial latencies are then summed for each satellite, and averaged over the constellation to determine the average delay for a piece of data to arrive from any given satellite on the network.

The second trial, modeling the On-Orbit Distribution architecture, as described in Algorithm \ref{alg:distributed} began with a random selection of satellites in the constellation as user/actuator nodes. For purposes of this study, we use the term actuator, but it is critical to note that all users of this received data may not actuate on the data. For every data source node on the network, the nearest on-orbit actuator is selected as the destination for the data. Several cases can result from this: 
\begin{enumerate}
    \item If the data are being collected at an actuator node, the latency is zero;
    \item  If the data source node can see a local actuator Node, the latency is the resulting propagation delay between the two assets;
    \item If the data source node does not have visibility to a actuator node, it will calculate a transmission path to the nearest available node. The net latency is then the sum of the delays over each `hop' in the transfer path.
\end{enumerate}

\subsection{Simulation Results}

\begin{figure*}[htp]
\centering
    \begin{subfigure}[b]{0.45\textwidth}
        \centering
        \includegraphics[width=\linewidth]{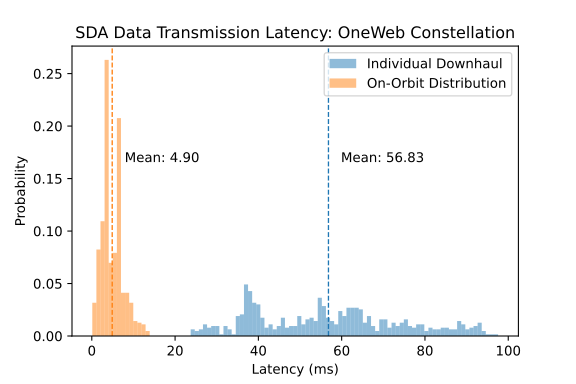}
        \caption{}
        \label{fig:demo_oneweb_latency}
    \end{subfigure}
    \hfill
    \begin{subfigure}[b]{0.45\textwidth}
        \centering
        \includegraphics[width=\linewidth]{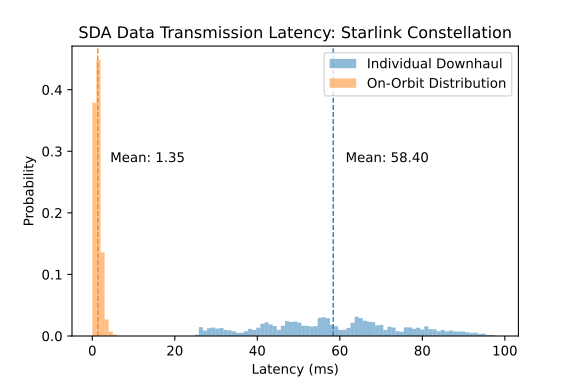}
        \caption{}
        \label{fig:demo_starlink_latency}
    \end{subfigure}
    \vskip\baselineskip
    \begin{subfigure}[b]{0.45\textwidth}
        \centering
        \includegraphics[width=\linewidth]{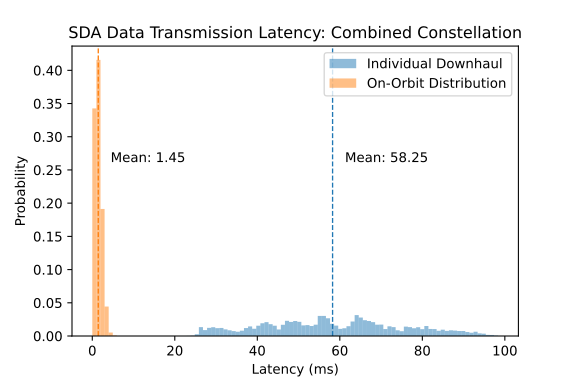}
        \caption{}
        \label{fig:demo_oneweb_starlink_combined_latency}
    \end{subfigure}
    \hfill
    \begin{subfigure}[b]{0.45\textwidth}
    \centering
        \includegraphics[width=\linewidth]{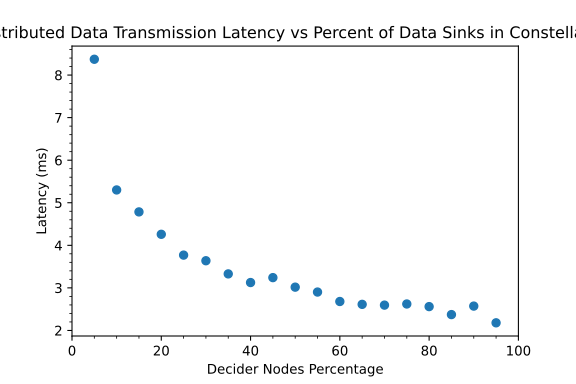}
        \caption{}\label{fig:distributed_data_sink}
    \end{subfigure}
    \caption{Simulation results for both network architecture for Starlink and OneWeb constellations: (a) Comparison on OneWeb constellation (b) Comparison on Starlink constellation (c) Comparison on combined Starlink and OneWeb constellation (d) Demonstration of the impact to latency of varying the number of the actuator nodes in OneWeb constellation.}
    \label{fig:combined_figures}
\end{figure*}

\begin{table}[!ht]
    \centering
    \caption{Simulation parameters for Starlink and OneWeb constellations.}
    \setlength{\tabcolsep}{7pt} 
    \renewcommand{\arraystretch}{1} 
    \begin{tabular}{|p{50pt}|p{80pt}|p{30pt}|p{30pt}|}
    \hline
       {Symbol} & {Description} & {Starlink} & {OneWeb}  \\ \hline 
        \( N_{total} \) & Total count of satellites & 5765 & 631    \\  
        \({S}\) & Active satellites set & & \\ 
        \( h \) & Altitude (km) & 550 & 1200   \\ 
        \( T \) & Orbital period (hrs) & 1.59 & 1.82    \\  
        \( D_{total} \) & Number of deciding nodes & 865 & 95    \\ 
        \({D}\) & Deciding satellites set & & \\
        \( G_{total} \) & Number of ground stations  & 13 & 13    \\ 
        \({G}\) & Ground stations set & & \\ 
        \( D \) & Centralized data allocator & & \\  
        \( M \) & Architecture mode & Both & Both  \\ \hline 
    \end{tabular}
    \label{tab:simulation_params}
\end{table}

Both architectures were tested with three sets of orbital assets; exclusively OneWeb and Starlink satellites, and a combined combination of both.  Table \ref{tab:simulation_params} details the comprehensive set of parameters used in the simulation of the two algorithms for both Starlink and OneWeb constellations.

Figures \ref{fig:demo_oneweb_latency} and \ref{fig:demo_starlink_latency} showcase the latency results for both individual downhaul and on-orbit distribution for the OneWeb and Starlink constellations, respectively. We observe that Starlink has better overall results, with an average latency of 1.35 ms compared to OneWeb's 4.50 ms for on-orbit distribution. However, the results for individual downhaul are very similar for both constellations, as the number of satellites does not significantly affect this configuration.

Figure \ref{fig:demo_oneweb_starlink_combined_latency} shows the results when we combine the OneWeb and Starlink constellations and evaluate the latency for both architectures. It is important to note that in this experiment, we achieved the lowest latency in the combined constellation at 1.45 ms (close to Starlink's individual performance). However, the combination did not improve the results for the individual downhaul, as the increased number of satellites from the combination did not have an impact.

Results were uniform across all three trials, with on-orbit distribution greatly outperforming individual downhaul in terms of latency. The increase in the number of satellites reduced the spread of on-orbit distribution latency and increased the gap between the distributed network's worst-case performance and the best-case performance of the downhaul architecture. The high latency of the latter is largely due to the greater distances needed to transmit from Earth's orbit to the ground, and then across terrestrial networks to the actuator node.

\begin{table*}[!htbp]
    \centering
    \caption{Relevant performance metrics and their impact on the two proposed reference architectures.}
    \label{tab:arch_comparison}
    \begin{tabularx}{\textwidth}{|l|X|X|}
        \toprule
        \textbf{Metric} & \textbf{Impact on Individual Downhaul Architecture} & \textbf{Impact on On-Orbit Distribution Architecture} \\
        \midrule
        Scalability & Ground infrastructure is a potential bottleneck for managing a growing fleet of satellites and increasingly sophisticated data collection. Both ground and space assets on the network must expand for consistent growth. The interdependency of LEO assets scales well with a growing constellation size. The network is easily augmented with MEO assets. & The network's scale is dependent on the number and distribution of member nodes. \\
        \midrule
        Reliability and Redundancy & When a centralized actuator is attacked, the network is disabled. & When one or several network fragments are compromised, others can operate as normal. \\
        \midrule
        Insight Quality & Additional terrestrial resources can be allocated to actuation, enabling more performant analysis. & Limited resources for on-orbit actuation may impact the quality and depth of insights generated by the network. \\
        \midrule
        Coverage & The network's coverage is dependent on the number and distribution of member nodes. & The network's coverage is dependent on the number and distribution of member nodes. \\
        \midrule
        Complexity & Low, when compared to other space-based wireless networks. & High, especially considering the variability in network nodes and operational challenges. \\
        \midrule
        Cost & Low, relies largely on existing infrastructure and capabilities. & Higher costs, associated with scalability, reliable high-throughput satellite cross-linking, and additional development costs for providing actuator capabilities to nodes. \\
        \midrule
        Security & Network is exposed to the much more prevalent threat to ground infrastructure, expanding its attack surface. & Completely on-orbit distribution necessitates a more sophisticated attack to successfully interdict or interrupt service. Increased complexity of orbital actuator nodes is another area of focus for a potential attacker. \\
        \midrule
        Latency & High latency on the ground segment. & Low latency, further improved with an increasingly fragmented network, although there are diminishing returns with fragmentation. \\
        \bottomrule
    \end{tabularx}
\end{table*}

The next study aimed to determine the impact that the percent of total actuator nodes in the On-Orbit Distribution architecture has on latency. First, there are two trivial cases:

\begin{enumerate}
    \item In the case of no actuator nodes, there is infinite latency because there is no destination for data.
    \item  Conversely, if every node is an actuator, there is effectively zero latency in the network, with each node reacting to only its own stimuli. 
\end{enumerate}

Figure \ref{fig:distributed_data_sink} shows a demonstration of the impact on latency when varying the number of actuator nodes in the OneWeb constellation. By breaking the network into fractions based on a percentage of actuator nodes, we demonstrated that while increasing the number of actuator nodes leads to latency improvements, these improvements show diminishing returns after roughly half of the nodes gain actuator privileges. This indicates the need to establish a balance between actuator nodes and latency, considering the trade-offs, especially given the additional cost of equipping an orbital node with the decision capabilities necessary to actuate on data.

\subsection{Advantages and Disadvantages}

Having analyzed the two reference architectures, it's evident that we can find different advantages and disadvantages for each. To effectively compare and contract the architectures and make meaningful performance comparisons, we believe the following set of metrics for tradeoff is essential:
\begin{itemize}
    \item \textbf{Scalability:} The ability for the architecture to adapt and expand to larger volumes of nodes and traffic on any segment;
    \item \textbf{Reliability and Redundancy:} The ability for an architecture to function independent of isolated failures on the network;
    \item \textbf{Insight Quality:} The inherent ability for the network to procure valuable data at high resolution.
    \item \textbf{Coverage:} The ability for the network to consistently observe a large area;
    \item \textbf{Complexity:} The sophistication (at a computational, resource-consumption, or routing level) required for a network to constantly maintain peak performance;
    \item \textbf{Cost:} The price to implement and maintain the architecture;
    \item \textbf{Security:} The measures to prevent or remediate relevant threats to the network;
    \item \textbf{Latency:} The time differential necessary for the network to process and respond to new data.
\end{itemize}

In Table \ref{tab:arch_comparison}, we compare the two proposed reference architectures on the basis of each metric described above. This analysis has led us to believe that while a distributed architecture entails a higher cost and increased complexity to maintain, these downsides will be mitigated with the advancement and commodification of LEO satellite infrastructure. Conversely, a distributed architecture presents a highly scalable and robust SDA sensing paradigm while operating at a lower latency to provide rapid response to crucial data.


In our comparison between individual downhaul and on-orbit distribution network architectures, we observe distinct impacts across various metrics, with security emerging as a pivotal consideration. In an individual downhaul architecture, scalability is hindered by potential bottlenecks in ground infrastructure, while reliance on a centralized actuator makes the network vulnerable to disabling attacks. However, the ability to allocate additional terrestrial resources can enhance insight quality.

On the other hand, an on-orbit distributed architecture offers scalability advantages, with the network's scale dependent on the number and distribution of member nodes. Although complexity and cost are higher due to operational challenges and additional development costs for nodes' actuator capabilities, the network gains resilience through redundancy, allowing normal operation even with compromised fragments. Moreover, latency is significantly reduced, especially with an increasingly fragmented network.

Security emerges as a critical aspect where the on-orbit distributed architecture excels. The transition to a completely on-orbit distribution model necessitates a more sophisticated attack to successfully interdict or interrupt service, considering the increased complexity of orbital actuator nodes.

\section{Security Considerations for SDA Networks}
\label{sec}

A fundamental element of establishing a robust network is its security. Therefore, we assess how an adversary would approach disabling or disrupting an SDA network. We have assumed that non-kinetic, electromagnetic threats comprise the largest risk to the space segment of the proposed SDA network reference architectures \cite{falco2021security}. The following work is limited to an analysis of potential interdiction on the link layer, the radiofrequency, optical, or—in the ground segment—physical cable links supporting the network. This excludes attacks at the physical layer, which includes attempts to compromise signal transceiver hardware, as well as the network layer, which includes attacks on the supporting routing and application management software. 

Additionally, cyberattacks can attempt to compromise data through a set of means known as the CIA Triad \cite{congress2002federal}: \begin{inparaenum}[(a)]
            \item confidentiality: attacks that threaten access control on sensitive data 
            \item integrity: attacks that threaten the veracity of transmitted data 
            \item availability: attacks which threatens the ability to access data when and where it is necessary.
\end{inparaenum}

\subsection{Attack Pathways}

Among the most common threats to the physical link layer is data availability, and as such it will be the focus of this analysis (though some attacks will have residual consequences on integrity). Notably, ensuring integrity and confidentiality directly addresses concerns highlighted in \cite{8756904} regarding the accessibility and disruptive potential of cyber-ASATs on crucial space systems. 

The simplest case for a link-layer attack is to render a link infeasible, typically through signal jamming (in the radiofrequency case) or a physical obstruction of the link (for optical transmission). Note again that attacks to the transmitter and receiver hardware itself comprise a physical-layer attack. Jamming attacks are typically executed with a target to ensure the interference signal strength is high enough to drown out true signal. As such, the effects are localized and either limited to the data transmitter or receiver. In the case of an SDA network, an attack to the transmitter (a satellite) completely disables SDA data from entering the network. While highly effective, this sort of attack is more challenging for an adversary to execute due to the remoteness of a satellite in comparison to a ground-based receiver.

In addition to an outright denial of data, the flow of data through the network can also be delayed by a link-layer attack. This case occurs if the network has implemented measures for disruption-tolerant networking. In such a scenario, if the network detects that data is not flowing through a link, it may be rerouted through active links of the network to reach its destination. The consequence of rerouting is a delay, both due to increased transmission time of the alternative path, and the delays necessitated to constantly monitor and verify the presence of disrupted links.

\subsection{Security Strategies}

To address these threats, we propose security strategies that align with the proposed attack pathway, focusing on the key aspects of availability, residual risk of integrity and confidentiality. 

\subsubsection{Redundancy and Reliability (Availability)}

Inherent redundancy and reliability within the network configuration mitigate the risk of a single node compromise disabling the entire network. However, effective implementation requires redundant communication pathways, adaptive routing algorithms, and rapid response protocols. These measures ensure continuous data availability even during disruptions and attacks, maintaining operational continuity and accessibility.

\subsubsection{Threat Detection and Response (Integrity)}

The proposed low-latency distributed architecture's design is particularly conducive to requirements of real-time threat detection mechanisms that focus on the integrity of data transmission. Continuous monitoring of data streams, detection of unauthorized modifications leveraging signatures block cyber intrusions onboard \cite{bailey2022cybersecurity}, and prompt response protocols, especially for actuator nodes, ensure the trustworthiness and accuracy of transmitted data within the SDA network. 

\subsubsection{Network Node Variability (Confidentiality)}

Variability in network node numbers introduces scalability, coverage, complexity, and cost considerations. While it offers increased coverage, it also heightens challenges in access control and key management. Strong authentication protocols, encryption and access controls on nodes are essential for preserving data confidentiality, limiting access to sensitive information to authorized entities only. Given the criticality of SDA data, stringent confidentiality measures prevent potential security risks from unauthorized access or data compromise.

\section{Future Research Directions}
\label{Future Research}

The unique requirements of an SDA network is a nascent field of study that demands further investigation and development. While this paper begins to assess architectural considerations for robust and scalable networks, many questions remain about designing proficient SDA networks that encompass both political and technological fronts. Therefore, a research agenda should be published on this topic. Such research is needed to enhance the safety, sustainability and security of space operations. In this section, we look at two key avenues for future research: standardization and policy management, and technological advances. 

\subsection{Standardization and Policy Management}
Current standardization and policy development efforts for SDA are mostly led at the national level. The international efforts are pioneered by the United Nations Office for Outer Space Affairs (UNOOSA) that aims to develop guidelines and best practices for space operations, including debris mitigation and long-term sustainability. Other notable collaborative activities  include Space Data Association \cite{spacedataassoc} and Space Safety Coalition \cite{SpaceSafetyCoalition}. Although these efforts are promising, there is limited focus on such standardization activities relating to space domain awareness. We assert there is a need for harmonized standards to ensure consistency across multiple constellations. Open communication and data exchange between nations and commercial entities should be encouraged to promote data sharing and consistency relating to SDA. 

Overall, this research direction emerges as the critical one to maintain the safety, sustainability and security of the New Space, with the goal of creating universally accepted guidelines for space operations, including debris mitigation and collision avoidance.
Additional necessary actions for safety, sustainability, and security include increasing public awareness on this topic, creating initiatives to educate stakeholders about best practices and compliance with space safety standards.

\vspace{-11pt}

\subsection{Technological Advancements}

The developments on standardization and policy developments should be supported by technological developments. As the most critical aspect, high resilience and accuracy sensors and tracking technologies are required to monitor space objects. The use of artificial intelligence and data fusion techniques will be critical to process large volumes of space data that will be gathered by these newly deployed sensors, in an efficient and effective manner. 

In these deployments, graph consensus algorithms \cite{torning2022} can play a crucial role in enhancing the safety, security, and reliability of SDA by ensuring data integrity, improving fault tolerance, and facilitating  collaboration among data nodes. Such algorithms can enhance the resilience of SDA systems against individual node failures or malicious attacks by achieving consensus among multiple nodes. They can also be used to detect discrepancies in data, identifying potential security threats or sensor malfunctions. Integrating these algorithms into SDA systems can significantly help achieve more accurate and secure space operations.

The combined use of sensing and communication can also help improve the system efficacy in this manner, where the waveforms used for radar sensing can also send data simultaneously \cite{Sumen_2022}. Note that integrated sensing and communication (ISAC) is already accepted as a feature of the emerging 6G networks \cite{mourad2023integrating}.  Once debris or an object to be removed is sensed, the next step involves using, the technologies for capturing and removing this space debris. In addition, spacecraft should be designed  with features that minimize debris creation.

By enabling these technological advancements, the safety, sustainability, and security of SDA can be significantly enhanced, ensuring the long-term viability of space operations.

\section{Conclusion}
\label{Conclusion}
In this work, we have identified the development of SDA networks as a critical research area. While many enabling technologies, such as improved sensing, tasking, and routing systems are subject to considerably inquiry, the development of scalable and robust architectures that acknowledges critical challenges to space infrastructure is still required. In approaching this problem, we present the current state of SDA networks and highlight key challenges, including security. We then define, at a basic level, the core components of an SDA network and construct a framework of performance metrics to evaluate them. In doing so, we identify two broad architectural approaches to constructing an SDA network; a centralized approach and a distributed approach. We foresee centralized approaches dominating the SDA landscape in the near-term -- drawing from our evaluation metrics, a centralized approach relies on largely preexisting hardware and link infrastructure and is close to how governments may instrument acquisition of SDA data today. On the other hand, a distributed approach is more resilient to disruptions on the ground segment and scales well with increasing population of dedicated sensing satellites on-orbit. Both architectures present security concerns, the trade-off being between exposed attack surface on the ground segment, or more complex and sensitive routing and actuation software posing weaknesses on the space segment. In both cases, there is significant work to be done in developing, optimizing, and securing SDA Network architectures to ensure accurate knowledge of the orbital environment for years to come.

\section*{ACKNOWLEDGMENT}
This work was supported in part by the Tier 1 Canada
Research Chair program and the Natural Sciences and Engineering Research Council of Canada (NSERC) Discovery program.

\ifCLASSOPTIONcaptionsoff
  \newpage
\fi



%

\bibliographystyle{ieeetr}
\bibliography{bibtex/bib/x-arxiv}

\end{document}